\documentclass[letterpaper]{article} 
\usepackage{aaai25}  
\usepackage{times}  
\usepackage{helvet}  
\usepackage{courier}  
\usepackage[hyphens]{url}  
\usepackage{graphicx} 
\usepackage{natbib}  
\usepackage{caption} 
\usepackage{algorithm}
\usepackage{algorithmic}

\usepackage{newfloat}
\usepackage{listings}
\DeclareCaptionStyle{ruled}{labelfont=normalfont,labelsep=colon,strut=off} 
\floatstyle{ruled}
\newfloat{listing}{tb}{lst}{}
\floatname{listing}{Listing}
\usepackage{xcolor}
\usepackage{booktabs}

\title{AI-Powered Algorithm-Centric Quantum Processor Topology Design}
\author{
    Tian Li\textsuperscript{\rm 1},
    Xiao-Yue Xu\textsuperscript{\rm 1},
    Chen Ding\textsuperscript{\rm 1},
    Tian-Ci Tian\textsuperscript{\rm 1},
    Wei-You Liao\textsuperscript{\rm 1},\\
    Shuo Zhang\textsuperscript{\rm 1},
    He-Liang Huang\textsuperscript{\rm 1}\thanks{Corresponding author.} 
}
\affiliations{
    \textsuperscript{\rm 1}Henan Key Laboratory of Quantum infommation and Cryptography\\ 
    Zhengzhou, Henan, 450000, China. \\

    quanhhl@ustc.edu.cn(corresponding author)
}

\usepackage{bibentry}

\begin{document}

\maketitle

\begin{abstract}
Quantum computing promises to revolutionize various fields, yet the execution of quantum programs necessitates an effective compilation process. This involves strategically mapping quantum circuits onto the physical qubits of a quantum processor. The qubits' arrangement, or topology, is pivotal to the circuit's performance, a factor that often defies traditional heuristic or manual optimization methods due to its complexity. In this study, we introduce a novel approach leveraging reinforcement learning to dynamically tailor qubit topologies to the unique specifications of individual quantum circuits, guiding algorithm-driven quantum processor topology design for reducing the depth of mapped circuit, which is particularly critical for the output accuracy on noisy quantum processors. Our method marks a significant departure from previous methods that have been constrained to mapping circuits onto a fixed processor topology. Experiments demonstrate that we have achieved notable enhancements in circuit performance, with a minimum of 20\% reduction in circuit depth in 60\% of the cases examined, and a maximum enhancement of up to 46\%. Furthermore, the pronounced benefits of our approach in reducing circuit depth become increasingly evident as the scale of the quantum circuits increases, exhibiting the scalability of our method in terms of problem size. This work advances the co-design of quantum processor architecture and algorithm mapping, offering a promising avenue for future research and development in the field.
\end{abstract}

%
\begin{links}
    \link{Code}{https://github.com/qclab-quantum/Qtailor}
\end{links}

\section{Introduction}
\label{sec:introduction}
The advent of quantum computing marks a significant leap forward in computational capabilities \cite{Wu2021StrongQC,Nielsenbook,ShorFactorization,deutsch,shorAlgorithmsQuantumComputation1994,ZHU2022240}, offering the potential to outperform classical computing in solving specific intricate problems. The field has witnessed remarkable progress in recent decades \cite{QuantumSupremacy2019}, having now entered the Noisy Intermediate-Scale Quantum (NISQ) era\cite{huangNeartermQuantumComputing2023,huangSuperconductingQuantumComputing2020,Wu2021StrongQC}. Despite this progress, the realization of quantum advantage in practical applications, such as quantum machine learning \cite{biamonteQuantumMachineLearning2017a,qnn,liuHybridQuantumClassicalConvolutional2021b,hhlgan,hhlsvm} and quantum chemistry \cite{bauerQuantumAlgorithmsQuantum2020,caoQuantumChemistryAge2019}, remains a huge challenge. This is primarily attributed to the fragile nature of physical qubits on noisy quantum processors \cite{decoherence}, making it difficult to execute complex quantum algorithms reliably. A pivotal strategy to mitigate the impact of noise is to minimize the depth of quantum circuits, which underscores the imperative for an efficient quantum circuit compilation process. This process involves mapping theoretical quantum algorithms onto the physical qubits within a processor, taking into account the native gate set and the processor's topology, with the goal of achieving the shallowest possible circuit depth.
 
Previous works have predominantly concentrated on circuit mapping for fixed processor topologies \cite{zhangTimeoptimalQubitMapping2021,liTackling2019,willeMapping2019,zulehnerEfficientMethodologyMapping2018,QubitAllocation2018,Isomorphismmaping,MCTS,Bipmapping}, such as linear or two-dimensional grid configurations. The topology of a processor significantly influences the depth of the mapped circuits, with higher connectivity generally leading to shallower depths. However, for certain quantum algorithms, a one-size-fits-all topology is not always necessary; instead, the processor can be tailored to the algorithm for optimal performance. This raises an intriguing question: Can tailoring the topology to specific circuits lead to further reductions in circuit depth? If affirmative, what methodologies can be employed to effectively engineer such tailored topologies? Our work addresses these questions by harnessing machine learning to simultaneously optimize the quantum processor topology and map quantum circuits for the intended quantum algorithm, thereby significantly reducing the depth of the mapped circuits. We summarize our key contributions as follows: 

\begin{itemize}
\item We introduce \textit{Qtailor}, a novel framework that utilizes Reinforcement Learning (RL) to determine the optimal quantum processor topology for a specific quantum algorithm, within the constraints of a processor's connectivity. Experimental tests with small-scale circuits indicate that \textit{Qtailor} can significantly reduce circuit depth, with enhancements reaching up to 46\% over the current state-of-the-art model, \textit{Qiskit} \cite{QiskitGithub}. Moreover, as circuit size escalates, the advantages of \textit{Qtailor} become progressively pronounced.
\item We develop a reward-replay mechanism for training \textit{Qtailor}, which recycles rewards from previous actions to reduce the need for time-consuming action evaluations, thus enhancing training efficiency and making \textit{Qtailor} well-suited for addressing large-scale problems.
\item We propose a directed-force layout technique that significantly minimizes edge crossings in the designed quantum computing processor, reducing crosstalk noise and optimizing the layout for the grid-like architecture commonly found in modern superconducting quantum computers.
\end{itemize}  

\begin{figure*}[t]
    \centering
    \includegraphics[width=0.9\textwidth]{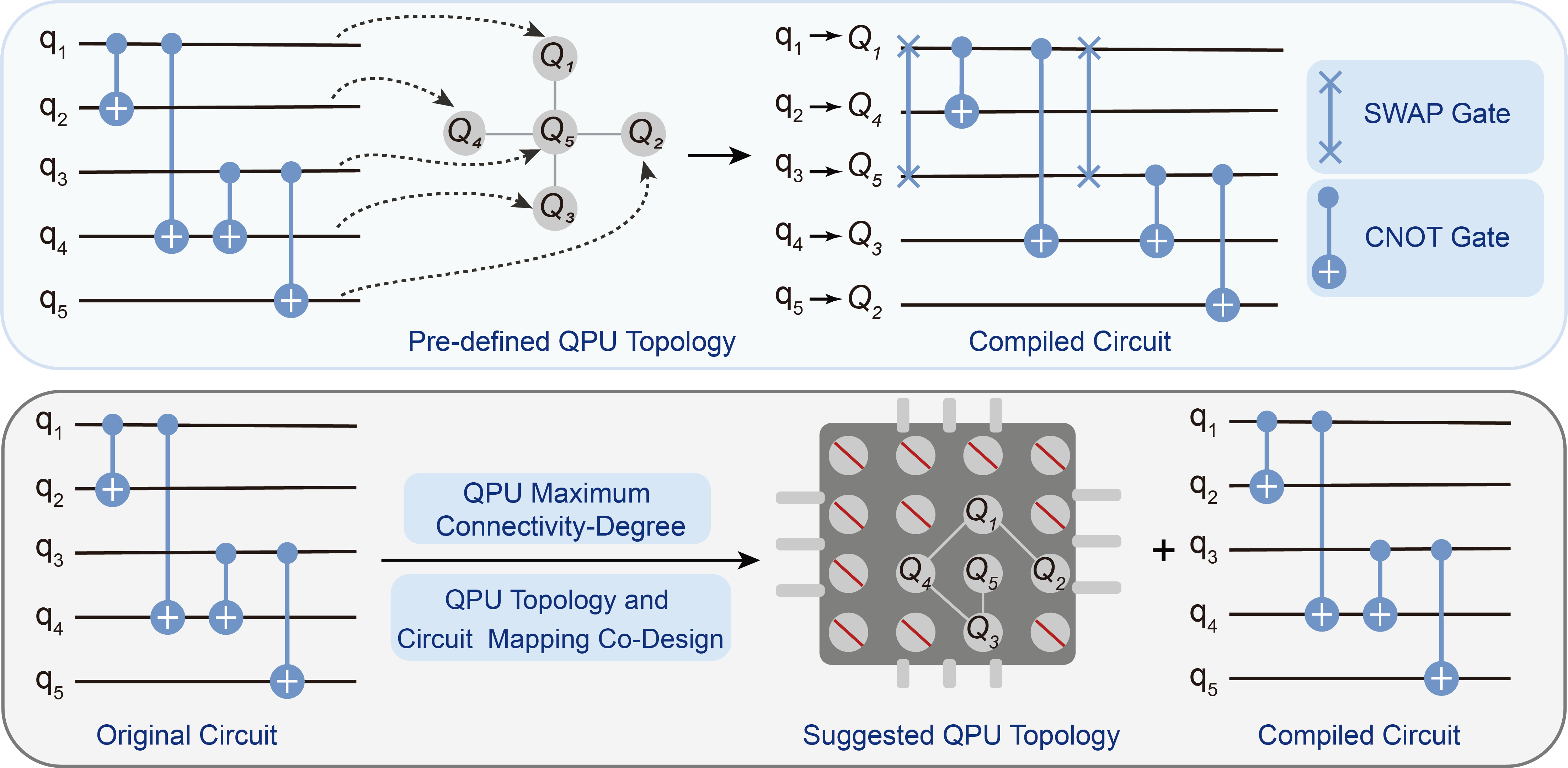} 
    \caption{Comparative  between the existing method and our method: \textbf{(top) }Prior research has focused on the development of mapping algorithms for predefined topologies, primarily aimed at allocating qubits in a way that satisfies connectivity requirements. The result of these methodologies is a fixed mapping protocol. \textbf{(bottom) } Our study employs Reinforcement Learning model to suggest an  topology that aligns with the circuit's characteristics under the limitations imposed by restricted connectivity. Subsequently, qubits are mapped in a sequential manner instead of a  complex mapping algorithms. }\label{fig:intro} 
    \end{figure*}
 
\section{RL for Processor Topology Design with Restricted Connections}

In the context of this study, we represent physical qubits and their connections using an undirected graph, with qubits as vertices (\textbf{v}) and their connections as edges (\textbf{e}). As elaborated in Appendix \ref{app:isomorphic}, our empirical evidence indicates that  variances between two graph  can lead to a significant divergence in circuit depth. It is, therefore, crucial to identify an optimal graph for circuits. Considering a set of $n$ vertices, the total number of possible edges is determined by the number of ways to select 2 vertices from $n$, hence, there are $2^{\frac{n(n-1)}{2}}$ potential graphs, with varying structures that markedly affects the circuit. Due to the exponential growth in potential  graphs with the number of vertices,  it is hard for heuristic search algorithms to traverse the extensive solution space. 
 
Reinforcement Learning \cite{silverMasteringGameGo2017a,intro_rl} is particularly  well-suited to this domain because its adaptive, incremental learning process,  a trade-off between exploration and exploitation, enabling effective exploration  in vast solution space. Consequently, our approach  leverages  RL to develop an agent that can dynamically guide topology design, moving beyond mere static memorization of solutions. Nevertheless, integrating  RL into  this domain presents  two primary challenges: \textbf{(a)} It requires the formulation of the topology design problem within the confines of a reinforcement learning framework. This involves developing novel action space and a reward function  that are compatible with reinforcement learning algorithms. \textbf{(b)} There is a critical need to improve the training efficiency of models designed for large-scale quantum circuits. This challenge arises from the inherent computational demands and the complexity of quantum circuits, which can significantly hinder the speed and effectiveness of model training. These challenges are addressed in Sections \ref{SystemOverview} and \ref{rr-ppo}, respectively.

\subsection{System Overview}\label{SystemOverview}
\begin{figure*}[t]
    \centering
    \includegraphics[width=0.99\linewidth]{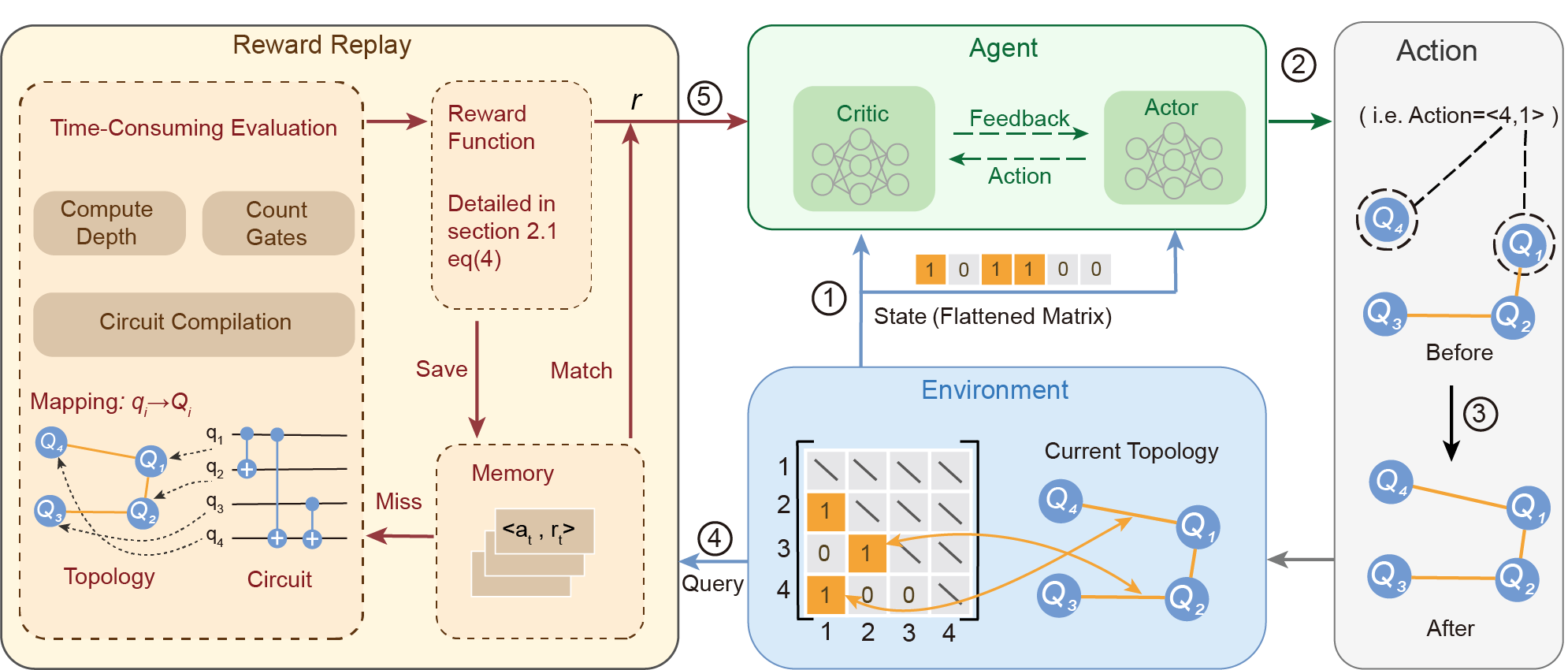}
    \caption{Overview of proposed \textit{Qtailor}: \textbf{ (1) }The agent acquires state from the environment; state are represented by a flattened matrix that denotes the current topology, where $M_{ij} = 1$ indicates that $Q_i$ and $Q_j$ are connected. \textbf{(2)} Subsequently, the agent outputs an action ($a$), that establish an connection between which two qubits. \textbf{(3)} The action is then applied to the topology. \textbf{(4)} Using the action as a key, we query the reward ($r$) from memory, which stores pairs of $\left \langle a,r \right \rangle $. If a match is found, the corresponding reward  will be directly provided to the agent, otherwise, an evaluation involving circuit compilation, computation of depth, and gates is conducted. The reward function is then applied based on the depth or gates, and this reward is stored in memory as a pair of $\left \langle a,r \right \rangle $. This process is referred to as reward replay, detailed in Section \ref{rr-ppo}. \textbf{(5)} Finally, The agent receives the reward and  continues to the subsequent iteration.}\label{fig:overview}
    \end{figure*}

Consider a Graph with $n$ vertices labeled $v_1, v_2, ..., v_n$, which correspond to $n$ qubits in a quantum  circuit. Let $E$ represent the complete set of potential edges, where each edge $e$, signifying a potential link between any two vertices (or qubits), belongs to this set, i.e., $e \subseteq E$. A Graph is thus defined as a subset of $E$ with each element of this subset representing an edge in the Graph. The depth of a circuit ($c$) mapped onto a Graph ($G$) can be computed using the function $f_{depth}(G, c)$. In our experimental setup, we utilized a famous quantum development kit \textit{Qiskit} \cite{QiskitGithub} to implement the computation of $f_{depth}(G, c)$. The objective is to identify a  $G$  to minimizes the circuit depth while adhering to the maximum connectivity degree constraint for its vertices, formally stated as:
\begin{equation}
\begin{array}{l}
 \min f(G, c),  G \subseteq E\\
  s.t. \;  degree(v) \leq 4  \; 
\end{array}
\end{equation}
In the context of the RL framework, we conceptualize  $G$ as the \textbf{environment}, wherein each action executed by the agent can modify $G$, propelling $G$ towards the desired state. The  $G$ is represented by an adjacency matrix, denoted by $M$, in which the presence of an edge between vertices  $v_i$ and $v_j$ is indicated by $M_{ij} = 1 $. It is important to note that Graph $G$ is undirected; consequently, an edge constitutes an unordered vertex pair, rendering $e\left \langle v_i, v_j \right \rangle $ and $e\left \langle v_j, v_i \right \rangle $ as equivalent representations of the same edge. In light of this symmetry, we use only the lower triangular portion of $M$ where $ i < j $, effectively reducing the state space by over half. As illustrated in Figure \ref{fig:overview}, matrix $M$ is subsequently flattened into a one-dimensional array to serve as the \textbf{state} (input) for the agent. The \textbf{agent} employed in our study is  borrowed from the Proximal Policy Optimization (PPO) as detailed in \cite{ppoOpenai}.
 
The extensive search space inherent in graphs makes it challenging for an agent to directly identify an optimal graph, especially during the initial stages when both the agent's policy and output are nearly random. This randomness results in sparse positive rewards, complicating the model's convergence. To address this, we simplify the huge action space into incremental steps. At each step, the agent only decide whether to build an edge between two specified vertices, consequently, the \textbf{action} can be denoted by $\left \langle v_i, v_j \right \rangle $. Given $n$ vertices the action space is effectively reduced to $\frac{n(n-1)}{2}$, a significant contraction from the entire Graph's search space. Reward is quantified as a scalar value  indicates the improvement of circuit depth (the lower the better) by action. An efficient reward function is  detailed in Section \ref{sec:rewardFunction}. Furthermore, to enhance the efficiency of training, a \textbf{Reward-Replay} method is introduced in Section \ref{rr-ppo}.
   
\subsubsection{Reward Function }\label{sec:rewardFunction}
Our objective is for the reward mechanism to guide the agent towards optimizing the topology towards an ideal state. We start by denoting the initial depth of the circuit as $D_0$, During the $i-th$  iteration,  the change in depth relative to the previous iteration is represented by $\Delta{(D_{i-1},D_i)} = D_{i} - D_{i-1} $, Our objective is to ensure that this change, $\Delta{(D_{i-1},D_i)}$, is negative, indicating a reduction in depth. Moreover, we strive for the depth at any iteration ${D_i}$ to be lower than the initial depth $D_0$, highlighting a consistent trend towards optimization. Consequently, the reward is computed based on both $\Delta{(D_0,D_i)}$ and $\Delta{(D_{i-1},D_i)}$. At any given time $t$, we assess  depth change $\Delta$ from step $t-1$ and the initial step to step $t$ respectively:
\begin{equation}
	\Delta D_{t \to 0}=\frac{D_t-D_0}{D_0}
\end{equation}
\begin{equation}
	\Delta D_{t \to t-1}=\frac{D_t-D_{t-1}}{D_{t-1}}
\end{equation}
According to Eq. (1) and Eq. (2), we formulate the reward as follows:
\begin{equation}
	r=\left\{\begin{array}{r}
		\left(\left(1+\Delta_{t \to 0}\right)^2-1\right)\left|1+\Delta_{t \to t-1}\right|, \Delta_{t \to 0} < 0 \\
		\\
		-\left(\left(1-\Delta_{t \to 0}\right)^2-1\right)\left|1-\Delta_{t \to t-1}\right|, \Delta_{t \to 0} \geq 0
	\end{array}\right.
\end{equation}
Our designed reward function incorporates both the initial state and the state from the preceding step, directing the agent to modify the topology at each step towards reducing the circuit depth. This ensures that the cumulative adjustments are aligned with the desired direction.

\subsection{Reward-Replay Proximal Policy Optimization}
\label{rr-ppo}
The reward function and action space operates independently of RL algorithms, enabling a variety of RL to tackle the mapping problem. Among these, Proximal Policy Optimization (PPO) \cite{ppoOpenai} is preferred due to its benefits in promoting stable training processes and efficient exploration strategies. Nonetheless, there still a significant obstacle that existing RL algorithms struggle to overcome: the evaluation of the action ( i.e. assigning rewards to actions) at each training step relies on computing the circuit's depth using $f_{depth}$, a process that can take several seconds to minutes. This evaluation process significantly reduces the training efficiency, consuming approximately 68\% of the total training time. This highlights   the crucial need to reduce this duration to improve training efficiency.

Typically, the evaluation associated with a particular action depends on the current state ($s$), indicating that identical actions may yield divergent outcomes across different states. This concept is encapsulated within the reward function $r(s,a)$. A significant challenge emerges due to the extensive array of potential states, making the storage of every possible state-action pair $\left \langle s, a \right \rangle $ impractical. However, our experiments have yielded a notable observation: across numerous distinct pairs $(s_p, a)$ and $(s_q, a)$ where $ s_p \neq s_q$,the results demonstrate a notable similarity in outcomes. In light of this observation, we simplify our model by adopting the assumption that $(a|s_1) \approx (a|s_2) \ldots \approx  (a|s_m)$, thereby enabling us to approximate the reward function for a state-action pair $r(s,a)$ as simply $r(a)$. This simplification is rationalized by the premise that if an edge significantly contributes to minimizing the circuit's depth by enabling the requisite connections, then, regardless of the current state (i.e., the graph), this edge consistently ensures the necessary connectivity, thus uniformly facilitating a reduction in the circuit's depth. Building on this approximation, we introduce the Reward-Replay method, which involves recycling the reward associated with an $a$ (represented as the pair $\left \langle r,a \right \rangle $) rather than $\left \langle r,a,s \right \rangle $, where the latter signifies a nearly infinite array of possibilities. This strategy allows for the immediate retrieval of rewards from memory upon any subsequent execution of the action, eliminating the need for repetitive reward calculations and consequently diminishing computational demands.

Utilizing  $r(a)$ to approximate the $r(a,s)$ may introduce inaccuracy. To counteract the potential accumulation of errors during training, a replay threshold is implemented for retaining rewards in the buffer. For instance, setting a threshold of 2 means that the reward associated with action will be discarded after its second use. The balance between training efficiency and performance can be adjusted by modifying the replay threshold. Details regarding the hyper-parameters and the pseudo-code for the Reward-Replay Proximal Policy Optimization (RR-PPO) are provided  in the Appendix \ref{app:ImplementationDetails}. It is important to emphasize that the framework we propose is not dependent on a specific reinforcement learning algorithm. Additionally, it does not require the use of a particular quantum software as a backend for processing quantum circuits.

\section{Topology Adjustment for Industrial Manufacturing}
\label{sec:layout}

Reinforcement learning offers promising   topology for circuits, however, translating these topologies into practical, industrially viable implementations encounters significant challenges: \textbf{(a)} Electromagnetic interference between crossing wires can cause a cross-talk\cite{FrequencyAwareCompilation} problem \cite{DetectingCrosstalkErrors2020}, leading to decoherence and low fidelity in quantum operations, necessitating a topology design that minimizes wire crossovers to alleviate such problems. \textbf{(b)} Over past decades, the quantum processor architecture based on superconducting qubits has become the leading candidate platform \cite{huangSuperconductingQuantumComputing2020}. Given the predominant grid-like structure of these  superconducting processor topologies \cite{IBM2020,zulehnerEfficientMethodologyMapping2018}, our goal is to maintain this uniform grid pattern. This approach aims to ensure compatibility with existing technologies, thereby facilitating a smoother integration of reinforcement learning-derived topologies into the current framework of quantum computing. 

\begin{figure}[tb]
	\centering
	\includegraphics[width=0.8\linewidth]{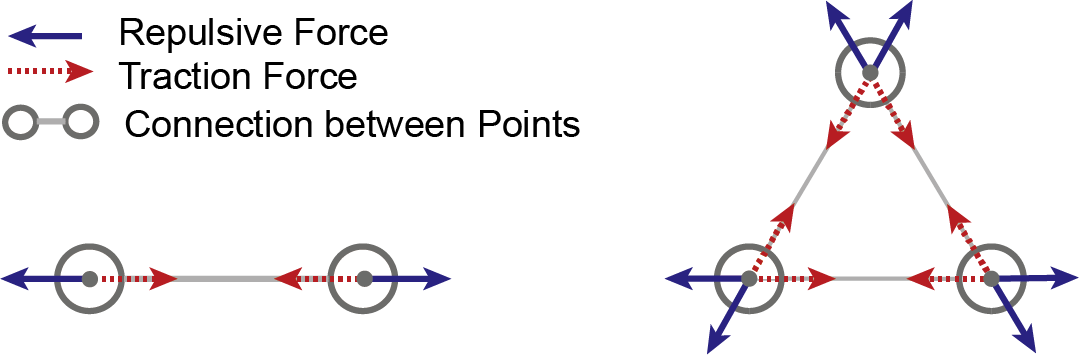}
	\caption{The forces acting between two and three vertex, each vertex undergoes both repulsive and attractive forces from the other vertex, with these forces counterbalancing each other to maintain the stability of vertex positions.}\label{force}
\end{figure}

\begin{figure}[tb]
	\centering
	\includegraphics[width=0.999\linewidth]{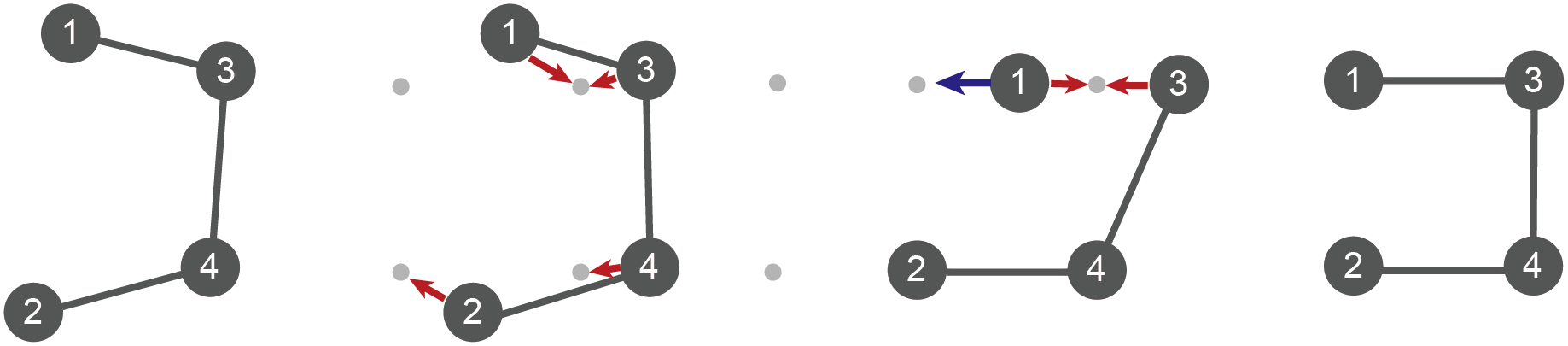}
    \makebox[\linewidth]{
        \makebox[0.249\linewidth]{(a)} \hfill
        \makebox[0.249\linewidth]{(b)} \hfill
        \makebox[0.249\linewidth]{(c)}\hfill
        \makebox[0.249\linewidth]{(d)}
    } 
	\caption{Workflow of the Force-directed Grid Layout: \textbf{(a)} Acquire an initial force-directed layout using existing method \cite{forceLayout}. \textbf{(b)} Initialization the grid point around the point, the point are  attracted to the nearest surrounding grid point, vertexes determine  the nearest grid point  via Euclidean distance, followed by occupation. \textbf{(c)} In instances where ${v}_{1}$ and ${v}_3$ share the closest grid point, they both move to that point. However, an increase in the repulsive force between them as they draw closer ensures that they are effectively repelled. \textbf{(d)} Upon occupation of a grid point by a point, the nearest unoccupied grid points is selected as the new target for other points.}\label{layout_process}
\end{figure}

In this study, we adopt the Force-directed Layout method \cite{forceLayout} to minimize edge crossings by emulating physical forces, specifically repulsion and attraction. As depicted in figure \ref{force}, vertexes are treated as charged particles, and edges are treated as springs, each vertex experiences two types of forces: \textbf{(a)} Repulsive force defined as $F_{repulsive} = \frac{k_1}{r^2} $, $k_1$ is a constant, $r$ is the distance between vertexes; \textbf{(b)} Attractive force: defined as $ F_{traction} = k_{2}\Delta{x} $, $k_{2}$ is a constant and $\Delta{x}$ is the value of the increase of the  distance between vertexes  compared to the initial distance.
The resultant force acting on each vertex is determined by the sum of all repulsive and attractive forces. Subsequently, the vertex adjusts its position in response to this resultant force. This adjustment process is iteratively repeated until the system reaches equilibrium, as detailed in Appendix \ref{appendixForce}. Equilibrium is characterized by a state of low entropy, leading to minimal edge crossings. While the layout method does not guarantee an absolute minimum number of crossings, its cost-effectiveness allows for multiple executions to identify the most optimal result. However, this method faces challenges in achieving a grid-like arrangement of vertices. To address this issue, we have incorporated a grid component into the existing force-directed method. This additional component directs vertices towards the nearest grid points, ensuring that vertices settle into a coherent grid pattern. The workflow of this component is illustrated in Figure \ref{layout_process}.

\section{Experiments}\label{sec:Experiments}
In this section, we present the experimental design aimed at addressing the following research questions: \textbf{(RQ1)} How does the performance of \textit{QTailor} compare to the state-of-the-art approach, and what are the reasons for the observed performance differences? \textbf{(RQ2)} How effective is  \textit{QTailor} as the circuit scales up? \textbf{(RQ3)} How does our proposed Reward-Replay impact the efficiency of PPO?

\subsection{Experiment Setup}
\textbf{Benchmarks and Compared Method .}
Building upon the foundation established by previous research, we have selected MQT Bench \cite{mqtbench} as our benchmark toolkit. This toolkit offers a diverse array of platform-independent circuits tailored for various quantum computing tasks. To monitor the training process and evaluate efficiency, we utilized TensorBoard \cite{tensorflow2015-whitepaper} 
Our experiments were conducted using \textit{Qiskit} \cite{QiskitGithub} as the backend framework, a renowned quantum computing development kit provided by IBM. Notably, \textit{Qiskit} incorporates Sabre \cite{liTackling2019} as its mapping algorithm, which is recognized as the state-of-the-art. Sabre utilizes a sophisticated mapping algorithm that operates on a SWAP-based Bidirectional heuristic search approach.

For our study, Sabre was chosen to ensure a fair and intuitive comparison. By employing Sabre, we maintain consistency across all phases of our experiment, with the exception of the mapping phase. This approach allows us to utilize the same configurations for post-mapping optimizations and circuit depth calculations, thereby ensuring a standardized and controlled environment for our comparative analysis.

\textbf{Evaluation Protocol.}
For the Sabre algorithm, a topology consisting of a square grid  with 100 nodes ($10 \times 10 $) is provided, where nodes having a connectivity degree of 4. For a fair comparison, the maximum connectivity degree for the \textit{QTailor's} topology is similarly constrained to 4. Given that the calculation of circuit depth involves randomness, we conducted the calculation three times and took the average to mitigate the effects of this randomness. The detailed hyper-parameters are given in Appendix \ref{app:ImplementationDetails}. 

\subsection{Performance Evaluation (RQ1)}
\label{RQ1}

\begin{figure}[tb]
	\centering 
	\includegraphics[width=1.0\linewidth]{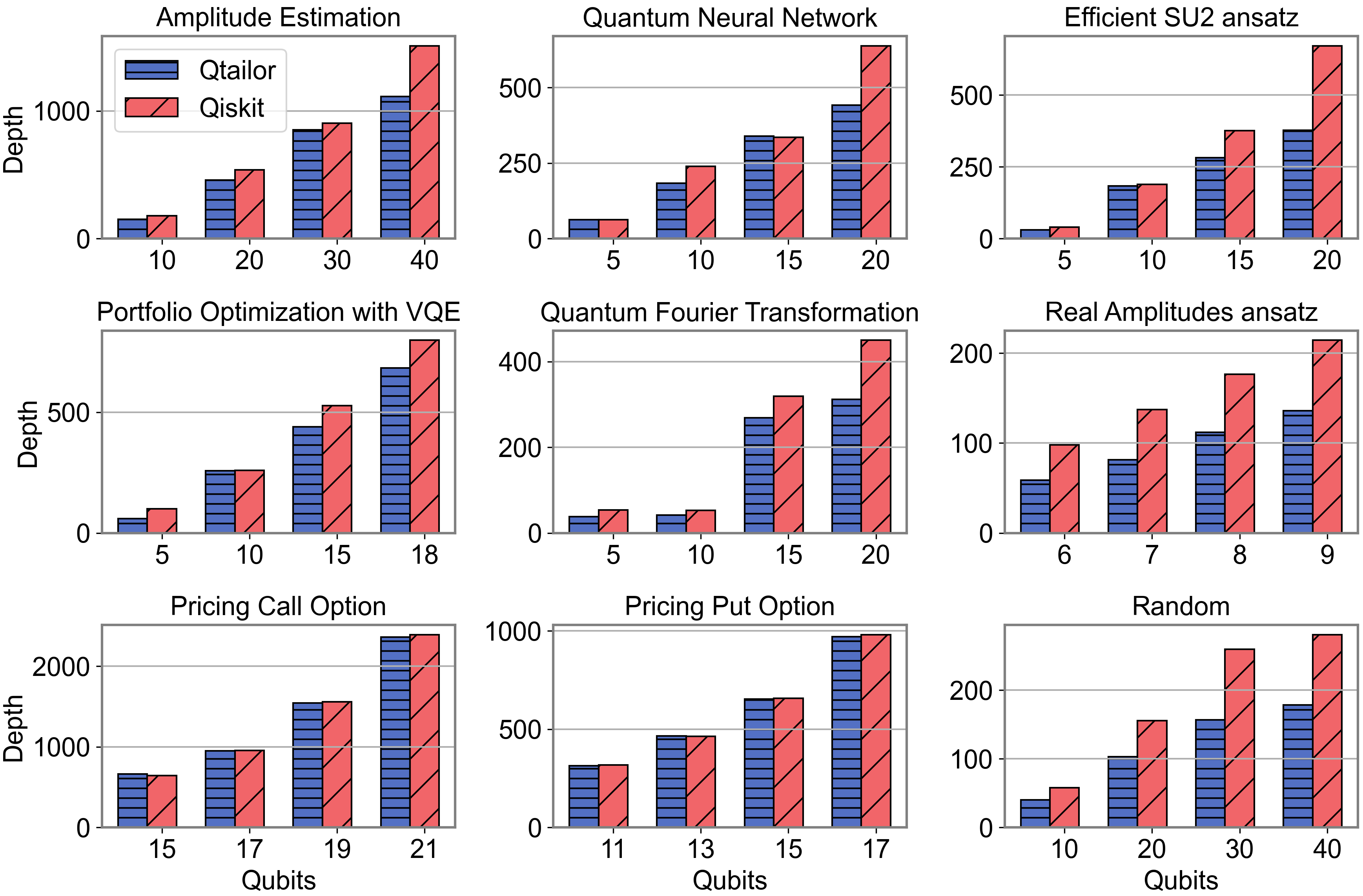}
	\caption{A comparative evaluation involving \textit{QTailor} and \textit{Qiskit}. The x-axis represents the circuit size quantified by the number of qubits, while the y-axis denotes the circuit depth after the mapping process.} \label{figure6}
\end{figure}

In this section, we compare the depth, fidelity, and total gates of circuits driven by \textit{QTailor} and \textit{Qiskit}.  The circuits therein is from public dataset. Our analysis indicates that \textit{QTailor} yields better outcomes than \textit{Qiskit} across a majority of tasks, reducing circuit depth ranging from 5\% to 46\% as presented in Figure \ref{figure6}, over 60\% of the samples exhibit a reduction in circuit depth by 20\%. The results shows that the topology suggest by \textit{QTailor} for circuits can  also significantly reduce total gates of compiled circuits, which can mitigate the accumulation of quantum errors, leading to higher overall fidelity. Our method can be easily optimized for different metrics by making a minor adjustment to the reward function. Specifically, we can replace the parameter depth (denoted by $\Delta{D}$) with the number of gates. Similar to depth, the number of gates should be minimized. Conversely, if the metric in question should be maximized, we need to use its negative value in the reward function.
 
The statistics presented in Table \ref{Idle_Ratio} provide insight into the factors contributing to the reduced depth of the line. The idle ratio, which represents the proportion of time a qubit remains inactive in gate operations, is calculated using the following formula:
\begin{equation}
	Idle = 1-(\frac{gates}{qubits \times  depth})
\end{equation}
  
\begin{figure}[t]  
    \centering
    \includegraphics[width=\columnwidth]{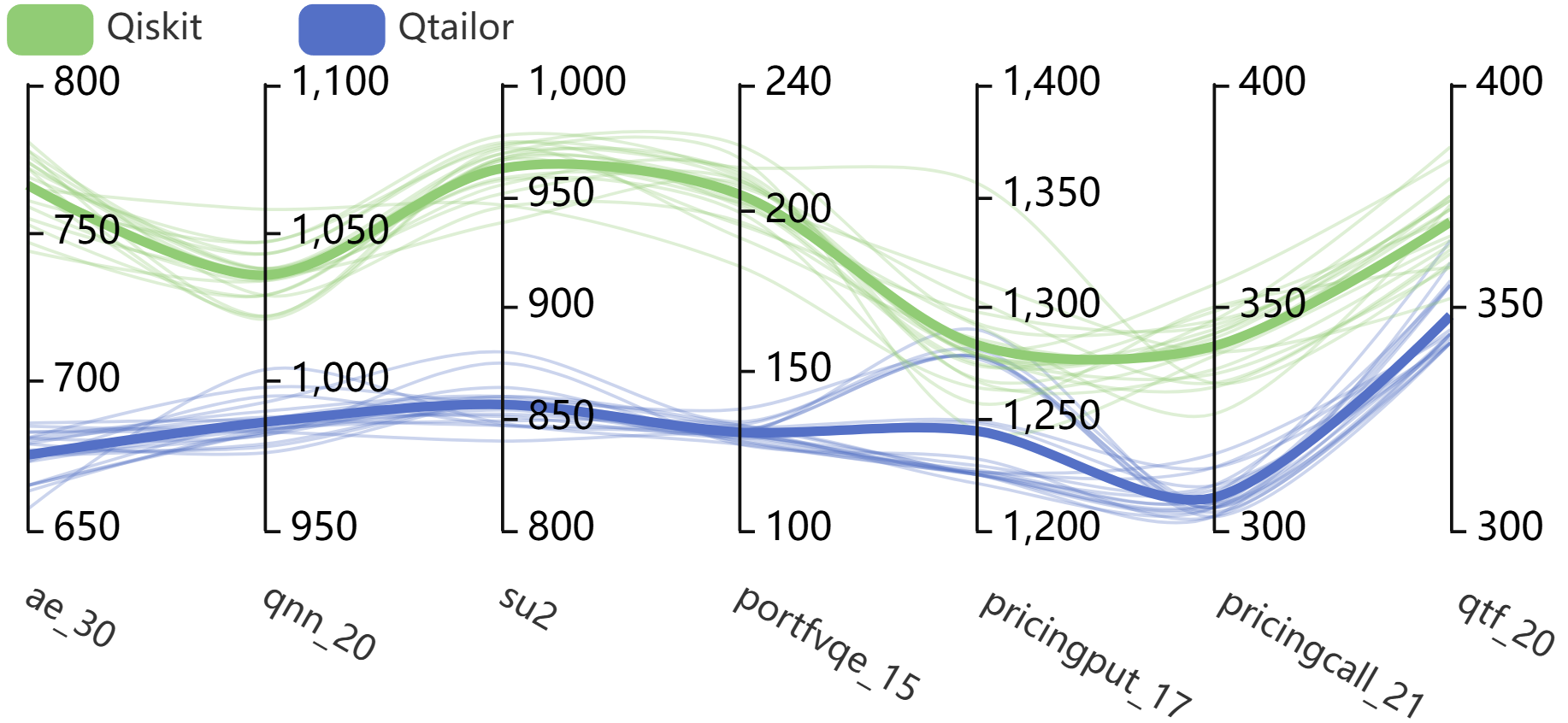} 
    \caption{Statistics on the number of gates. Each curve represents a  single experiment conducted across all circuits, with the bold curve indicating the mean value. Our approach reduced the total number of gates by 4.78\% to 36.39\%.}\label{fig:gates} 
    \end{figure}

\begin{table}[t]
    \centering
        \setlength{\tabcolsep}{1mm}{
    \begin{tabular}{c*{2}{|*{2}{c}}|c}
        \toprule
        & \multicolumn{2}{c|}{\textit{Qiskit}} & \multicolumn{2}{c|}{\textit{Qtailor}} 
                        \\
                        Circuit &  Depth  &  Idle(\%) &
                        Depth  &  Idle(\%) &Reduction(\%)
                        \\
        \hline 
        qnn\_5  &      92 &  76.52  &	63&	   66.70 & 31.52$\downarrow$\\
        qnn\_10  &	191 &	75.29  &	184&	76.47 & 3.66$\downarrow$\\
        qnn\_15  &	483 &	84.89  &	330 &	77.52 & 31.68$\downarrow$ \\
        qnn\_20  &	610 &	83.67  &	445 &	74.04 & 27.05$\downarrow$\\
    vqe\_10   & 257 & 74.09   & 253 & 67.39 & 9.04$\downarrow$ \\
    vqe\_15  & 507 & 79.20  & 428 & 69.66 & 12.04$\downarrow$ \\
    ae\_10  & 191 & 84.29   & 162  & 81.67  & 3.11 $\downarrow$ \\
    ae\_20  & 527 & 74.65    & 459  & 70.04  & 6.18 $\downarrow$ \\
    ae\_40  & 1353 & 58.47    & 1069  & 40.79  & 30.24 $\downarrow$ \\
    ansatz\_6   & 95 & 86.32    & 55  & 81.27  & 5.85 $\downarrow$ \\
    ansatz\_9 & 193 & 83.63   & 136  & 81.84  & 2.14 $\downarrow$ \\
    su2\_5  & 62 & 87.10    & 22  & 74.55  & 14.41 $\downarrow$ \\
    su2\_20  & 480 & 64.21    & 394  & 41.24  & 35.77 $\downarrow$ \\
    \bottomrule
    \end{tabular}
    }
    \caption{Statistics of  depth, and idle ratio on circuits. Results suggests that minor variations in idling rates can culminate in substantial differences in circuits depths. }
    \label{Idle_Ratio}
    \end{table} 
We observe that the idle ratio from \textit{Qtailor} is  lower than that of \textit{Qiskit}, with a maximum observed reduction of 31.68\%, regardless of whether \textit{Qtailor's} gate count is higher or lower than \textit{Qiskit's}. A lower idle ratio indicates enhanced parallel processing capabilities of the quantum processor and contributes to a reduction in circuit length.  If the focus shifts towards reducing the number of gates, a minor adjustment can be made to the reward function by replacing the depth parameter (denoted by $\Delta{D}$)  while other  components remain unchanged. The result are shown in Figure \ref{fig:gates}. 
     
\begin{figure}[tb]
    \centering
    \includegraphics[width=\columnwidth]{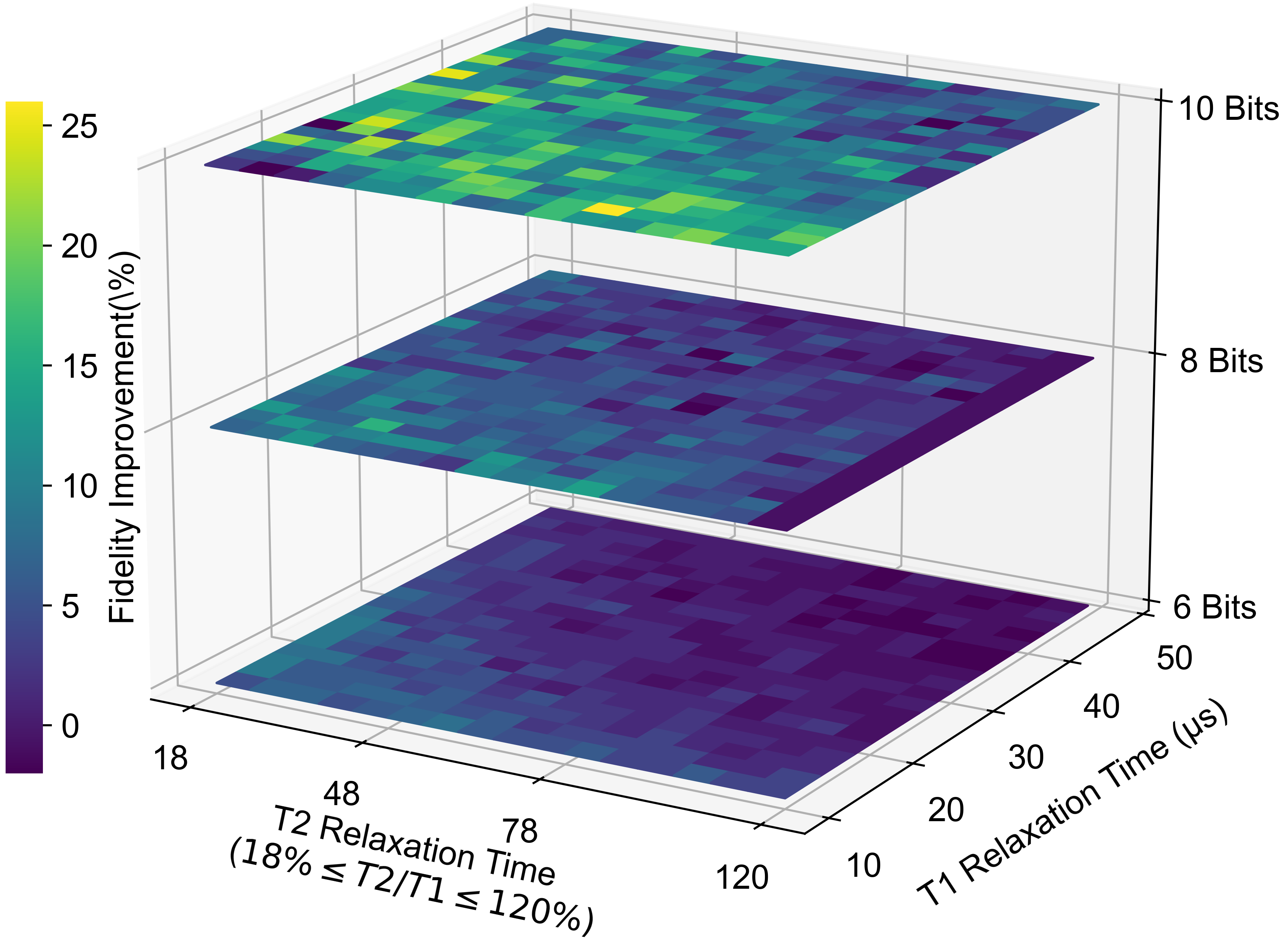}
    \caption{Fidelity improvement with gates reduction, the surfaces from top to bottom represent circuits of 6, 8, and 10 bits of Real Amplitudes ansatz task. }\label{fig:fidelity} 
\end{figure}

    The Quantum gates in circuits are applied to qubits to perform computations. Each gate takes a certain amount of time to execute, the gates must be executed within an operational window dictated by the qubits coherence times, such as T1 (relaxation time) and T2 (dephasing time). The T1 and T2  define the duration for which a qubit can maintain its state reliably, if operations exceed these time limits, qubits may lose their state, leading to errors. Fewer gates  results in shorter operation times, which is crucial for enhancing fidelity—a metric indicating the accuracy of a quantum operation or circuit performance.  The improvement in fidelity achieved through our method is depicted in Figure \ref{fig:fidelity}. We employ \textit{Qtailor} to optimize and reduce the quantum gate count in the circuit. Next, we evaluate the fidelity of the circuit using Equation \ref{eq:fpq}, with  $P_i$ and $Q_i$ denoting  the probabilities of the i-th quantum state for the ideal and noise-affected scenarios, respectively. The findings reveal that as the circuit size grows, the fidelity enhancement becomes more significant. In a 10-qubit circuit, there is an average fidelity improvement of 10.91\%, with a maximum enhancement of 26\%.
   \begin{equation}
    \label{eq:fpq}
   F(P,Q)=\left(\sum_i\sqrt{P(i)Q(i)}\right)^2
   \end{equation}
\subsection{Evaluation of Circuits for Scaling Up (RQ2)}\label{sec:RQ2}
\begin{figure}[tb]
	\centering
	\includegraphics[width=0.70\columnwidth]{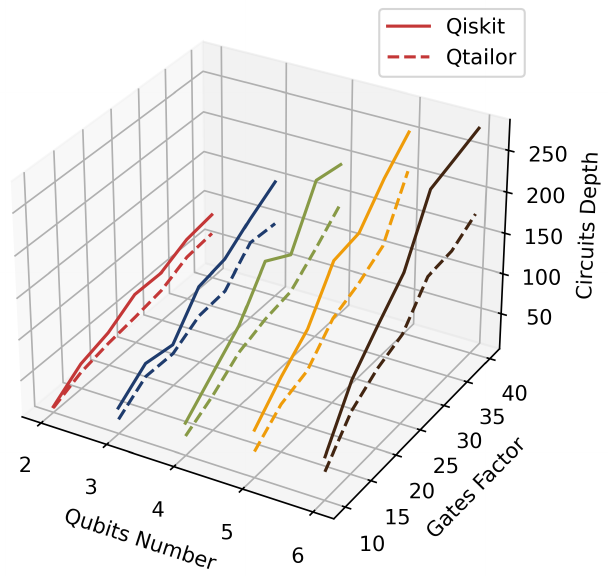}
	\caption{Depth of mapped circuits of varied sizes. The circuit's size is determined by both the number of qubits and gates. The number of gates is determined by multiplying the number of qubits by a factor, such as 2 on the axis indicating that the number of gates equals twice the number of qubits.
	}
	\label{RQ2} 
\end{figure}
Figure \ref{RQ2} shows how the depth changes with size of the circuits increase. In a comparative analysis focusing on the depth of quantum circuits comprising 6 qubits, \textit{Qtailor} consistently exhibits a reduced circuit depth when compared to \textit{Qiskit}. As the gate count increases from 60 ($6 \times 10$) to 210 ($6\times35$), the depth reduction achieved by \textit{Qtailor} compared to \textit{Qiskit} are \textbf{31.03\%, 31.52\%, 33.87\% , 39.11\%, 39.65\% and 40.74\%}, respectively. This trend indicating a systematic efficiency improvement in circuit depth with \textit{Qtailor} as circuits size increases. The capability to efficiently handle large-scale circuits is crucial for unlocking the full potential of quantum computing, thereby enabling the solution of problems that are intractable for classical computing. 
 
\subsection{Ablation Study (RQ3)}
\label{sec:RQ3}
\begin{figure}[t] \centering 
    \includegraphics[width=\columnwidth]{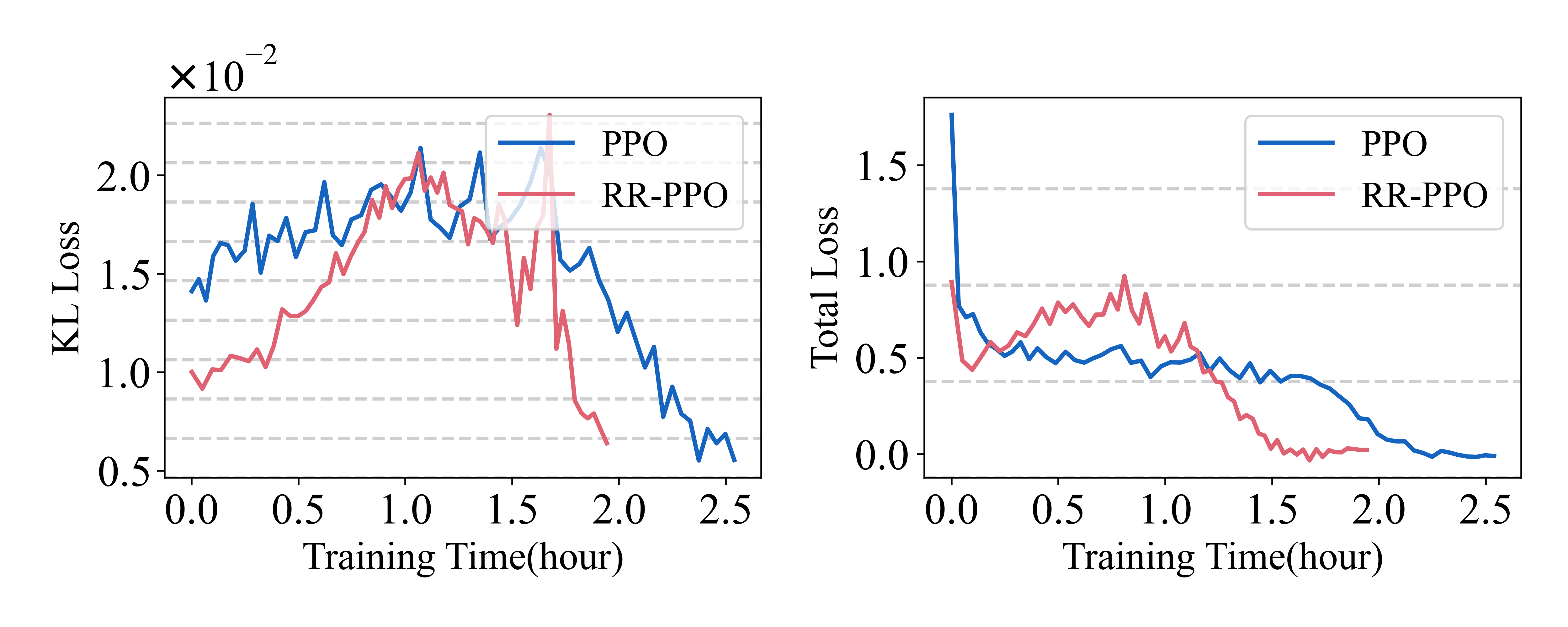}
    \caption{Comparison of the loss curves for PPO and  RR-PPO. After a period of training, RR-PPO requires less time for sampling , resulting in a decrease in the time consumed per iteration.} 
    \label{figure7} 
\end{figure}
\begin{figure}[t] \centering
	\includegraphics[width=\columnwidth]{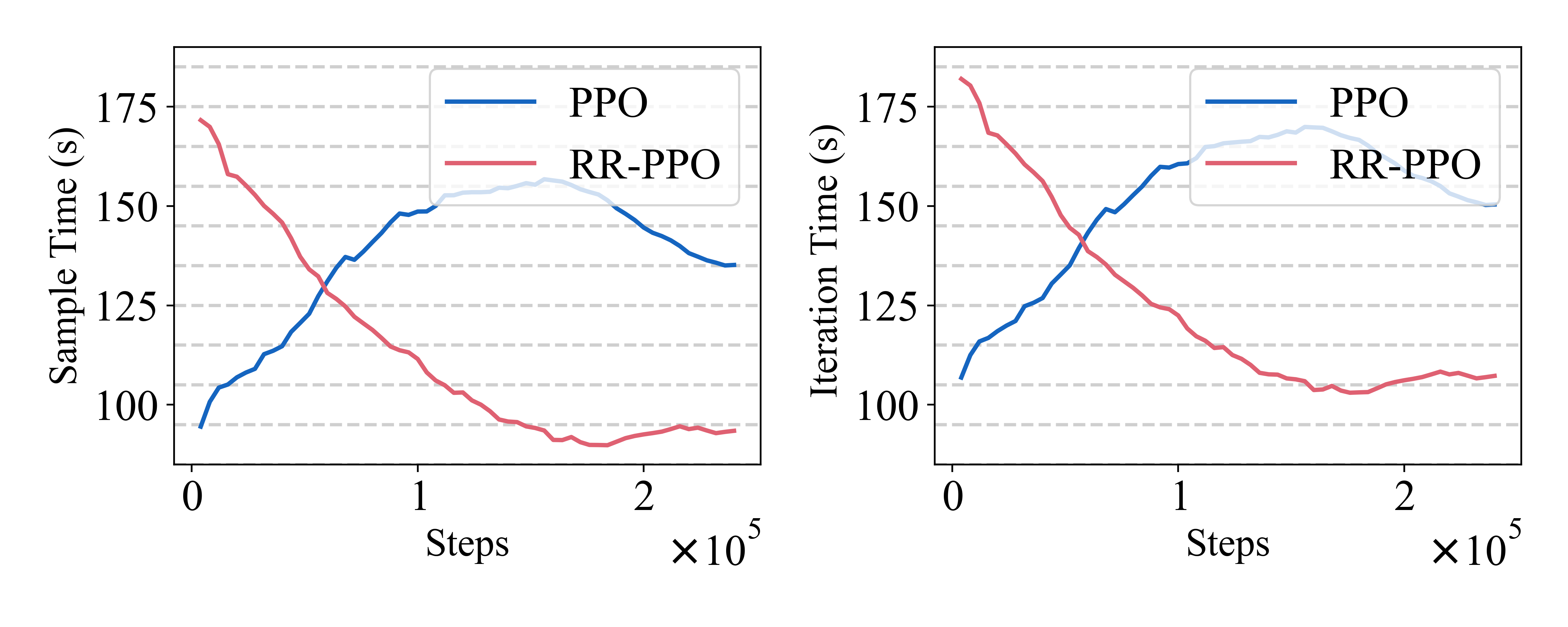}
    \caption{Time consuming. The time required to collect one sample \textbf{(left)} and to complete one iteration \textbf{(right)}. An iteration comprises sampling, gradient descent, and updating the policy. The results indicate a rapid decrease in the time expended by RR-PPO over time.} 
    \label{fig:figure9}
\end{figure}  
Figure \ref{figure7}  shows the Kullback-Leibler (KL) divergence loss and  total loss during the training. Besides the Reward-Replay module, the experiments are conducted under uniform conditions, including hardware and hyper-parameters. We observe that: 
\textbf{(left)} Our method shows a more fluctuating  curves for  both KL  and total loss, yet  reaches convergence faster by 24\% . This efficiency  can be attributed to the Reward-Replay module, which minimizes unnecessary evaluations on circuits.
\textbf{(right)} The losses curves eventually reach a nearly a near-identical value, because the forgetting mechanism avoids the  errors accumulation and thus has a limited effect  on  training accuracy.

Figure \ref{fig:figure9} provide insights into the faster training process of RR-PPO, The results shows that the time consumed by RR-PPO gradually decreases and stabilization  on both sampling and iteration phrase. Yet, the traditional PPO exhibits a contrasting trend. As training progressed, more reward can be replayed to avoid time-consuming evaluations. We also notice that the trends in the sampling time \textbf{(left)}  and iteration time \textbf{(right)} curves exhibit strong similarity. Additionally, it is noteworthy that the iteration time encompasses the sampling time, implying that  the total time to complete an iteration is mainly determined by the evaluation in sampling process.
\section{Discussion and Conclusion}
\label{sec:Discussion}
The proposed \textit{Qtailor}, a framework that pioneers the synergistic design of quantum processor topology and circuit mapping, offers several remarkable features: 1) Compared to state-of-the-art methods, \textit{Qtailor} significantly reduces the depth of mapped circuits, with a minimum of 20\% reduction observed in 60\% of the examined cases and a maximum enhancement reaching up to 46\%. This reduction is pivotal for high-fidelity quantum computing, as circuit depth is a critical factor affecting the fidelity of quantum operations. This work, therefore, highlights the transformative potential of machine learning techniques in advancing quantum computing by transcending the limitations of conventional compilation methods. 2) Our method demonstrates excellent scalability across problem sizes and training efficiency. As illustrated in Section \ref{sec:RQ2}, the advantages of our approach become more pronounced with larger problem sizes, indicating its effectiveness for complex quantum circuits. Furthermore, as detailed in Section \ref{sec:RQ2}, the reinforcement learning framework we designed, which encompasses problem modeling and a problem-oriented training strategy called reward-replay proximal policy optimization, has yielded significant improvements in training efficiency. 3) We have also integrated a force-grid layout technique to optimize the topology on the processor, enhancing compatibility with current manufacturing technologies and mitigating issues related to crossing connections. This integration is crucial for the practical realization of quantum processors and addresses a key challenge in the transition from theoretical designs to manufacturable devices.

In the current study, we have focused on the connectivity constraints of quantum processors, an essential factor in processor design. However, different quantum physical systems or varying fabrication processes for quantum processors may entail distinct constraints. While our work provides a preliminary framework, it sets the stage for future research that can explore these nuances more deeply. 

\appendix
\section{Appendix}
\subsection{Related work}
Circuit mapping involves the judicious allocation of qubits on a quantum processor which is known to be NP-complete \cite{Boixo2018}. One common approach is to reformulate the  mapping issue as an mathematic problem and utilize a software solver \cite{Determining2015,Greedy2018,ijcai2017p620,Venturelli_2018,bhattacharjeeDepthOptimalQuantumCircuit2017}. Wille et al. \cite{willeMapping2019} addressed the mapping problem by incorporating it into a Satisfiability Modulo Theory(SMT) framework \cite{de2009satisfiability} and they have successfully utilized the Z3 boolean satisfiability solver \cite{z3slover} 	 to to obtain precise results for mapping quantum circuits onto IBM QX architectures. Another proposed method, BIPMapping \cite{Bipmapping} by Nannicini et al. models the mapping problem as an Integer Linear Problem (ILP) \cite{schrijver1998theory} and and employs IBM CPLEX \cite{cplex} as a slover, the goal is to minimize a linear objective function subject to a set of linear constraints. BIPMapping consists three possible objective functions including error minimize, depth minimize and cross-talk minimize. Notably, BIPMapping outperforms Sabre in terms of reducing the number of CNOTs and exhibits higher fidelity.

A significant limitation of aforementioned methods is that a general solver is suffer from long runtimes and are only applicable to small-scale cases.As a solution, researchers have explored the use of search-based methods to find optimal mapping schemas \cite{zhangTimeoptimalQubitMapping2021, zhangOneQCompilationFramework2023,burgholzerLimitingSearchSpace2022,2023sanjiang,MCTS,sanjaing2020,Isomorphismmaping,microarchitecture2017, zhangDepthAwareSwapInsertion2020}. Siraichi et al. \cite{QubitAllocation2018} conducted a study on qubit mapping problems with IBM QX2 and QX3 processors, proposing a search algorithm based on dynamic programming to identify optimal solutions. However, this optimal algorithm requires exponential time and space for execution and is constrained to circuits with a maximum of 8 qubits.  In 2019, Siraichi et al. \cite{siraichiQubitAllocationCombination2019} further developed an improved algorithm and evaluated its performance on IBM Q20 Tokyo. They employed a Bounded Mapping Tree (BMT) to explore the complex landscape of qubit mapping by breaking the problem down into subgraph isomorphism and token swapping problems.
 
Zulehner et al. \cite{zulehnerEfficientMethodologyMapping2018}  employed $A^*$ Search algorithm \cite{astar} along with a heuristic cost function to optimize the arrangement of two-qubit gates in separate layers, similar to Sabre. The main objective was to minimize the overall distance between coupled qubits in each layer while simultaneously reducing the circuit's depth. This approach taking only a few minutes to execute on small-scale circuits. However, it requires the analysis of all possible combinations of concurrent SWAP gates, which result in an excessively long runtime.

IBM introduced Sabre \cite{liTackling2019} a state-of-the-art mapper with optimized heuristic search,Sabre utilizes the Floyd-Warshall algorithm to calculate the All Pairs Shortest Path and generate the matrix $D$. In this matrix, $D[i][j]$ represents the minimum number of swaps required to transfer a logical qubit from physical qubit $Q_i$ to $Q_j$. The computation complexity of this step is $O(N^3)$. Afterward, Sabre traverses the entire circuit to create a Directed Acyclic Graph (DAG) that visualizes gate dependencies, with a complexity of $O(g)$. The DAG is then divided  into independent layers, enabling concurrent execution of gates within each layer. 

Liu et al. \cite{liuNotAllSWAPs2022} proposed an optimization-aware algorithm  that utilizes a similar mapping strategy as Sabre. However, their algorithm calculates the reduction in CNOT gates for each SWAP candidate in the routing process and selects the best swap candidate based on a cost function. Experimental results demonstrate that their algorithm reduces the number of CNOT gates by an average of 21.30\% and the circuit depth by an average of 7.61\% compared to Sabre.

Another important perspective is that circuit mapping should consider noise, as real-world processors exhibit significant variability in the error rates of qubits and their connections \cite{IBM2020}. Tannu et al. \cite{tannuNotAllQubits2019} introduced a Variation-Aware Qubit Allocation (VQA) approach, which optimizes the movement and allocation of qubits towards stronger qubits and connection, thereby enhancing the reliability of NISQ systems by up to 2.5 times. Similarly, Murali et al. \cite{muraliNoiseAdaptiveCompilerMappings2019} conducted mappings on a real system using daily calibration data provided by IBM to prioritize qubit positioning, resulting in a reduced likelihood of communication (SWAP) errors. The results demonstrate that proper placement can lead to over 10 times improvement in the success rate of execution. However, this method suffers from the limitation that the initial qubit layout, which is optimal in terms of noise characteristics, may not be ideal for later qubit routing through swap mapping to the target topology. To overcome this issue,  Nation et al. proposed MAPOMATIC \cite{nationSuppressingQuantumCircuit2023} which involves remapping quantum circuits after qubit routing or post-compilation, to match low-noise subgraphs. They utilize output circuits generated with a SWAP-efficient layout and routines that are insensitive to device parameters,such as the the SABRE \cite{liTackling2019}, and then score each mapping result using heuristic cost functions derived from calibration data. Finally, the circuits are remapped to the lowest error subgraph before execution. The results demonstrate that using better qubit selection can recover, on average, nearly 40\% of missing fidelity.

While all of the aforementioned methods have produced notable outcomes. they have primarily focused on circuit mapping for fixed processor topologies. None of these previous studies have specifically customized the processor to the algorithm in order to achieve optimal performance.

\subsection{More Background}
\textbf{Qubits}: In the field of quantum computing, the qubit, or quantum bit, functions as the counterpart to the classical bit. Diverging from the classical bit's limitation to representing solely '1' or '0', a qubit can simultaneously exist in a coherent superposition of both states. This feature establishes it as a two-state quantum system that encapsulates the distinctive attributes of quantum mechanics \cite{Nielsenbook}. An exemplary illustration of this concept is the electron spin, characterized by its two potential states: spin-up and spin-down.

\textbf{Quantum Gates}: Quantum computing operations fundamentally rely on two primary categories of quantum gates. The first category includes single-qubit gates, which are unitary operations that can be conceptualized as rotations around the axis of the Bloch sphere, representing the state space of a single qubit \cite{Nielsenbook}. The second category consists of multi-qubit gates, which are crucial for executing more complex quantum computations. Notably, complex quantum operations can be decomposed into sequences of single-qubit gates—namely, H, S, T gates—and the two-qubit Controlled NOT (CNOT) gate. The CNOT gate, due to its critical role, will be the primary focus of this discussion. It operates on two qubits, designated as the control and the target. The principle of operation for the CNOT gate is simple: if the control qubit is in the state '1', it flips the state of the target qubit; if the control qubit is '0', the target qubit remains unchanged.

\textbf{Quantum Circuit}: A quantum circuit consists of a collection of qubits and a sequence of operations applied to these qubits. The representation of quantum circuits can be achieved through several methods. One such method employs the quantum assembly language known as OpenQASM, introduced by IBM \cite{crossOpenQASM2022}. Alternatively, circuit diagrams can be used, where qubits are depicted as horizontal lines and quantum operations are illustrated by various blocks placed along these lines.

\subsection{Comparison with Tket Compiler}
\textit{Qtailor} utilizes \textit{Qiskit} to assess circuits depth and gate count. However, it is important to note that \textit{Qtailor} is not limited to a specific compiler; rather, it serves as a framework that is compatible with multiple compilers.  The QPU topology suggested by \textit{Qtailor} is generic. To illustrate this flexibility, we performed a comparative analysis with \textit{Tket} \cite{tket}. Aside from switching compilation platforms from the \textit{Qiskit} to \textit{Tket}, no further  modifications were necessary. The topology used in the experiments remained consistent with those in Section 4, specifically a fixed grid-like topology and the suggested topology from the RL module. Consequently, there is no requirement to retrain the RL model when changing compilers.

\begin{figure}[t] \centering
    \includegraphics[width=0.8\columnwidth]{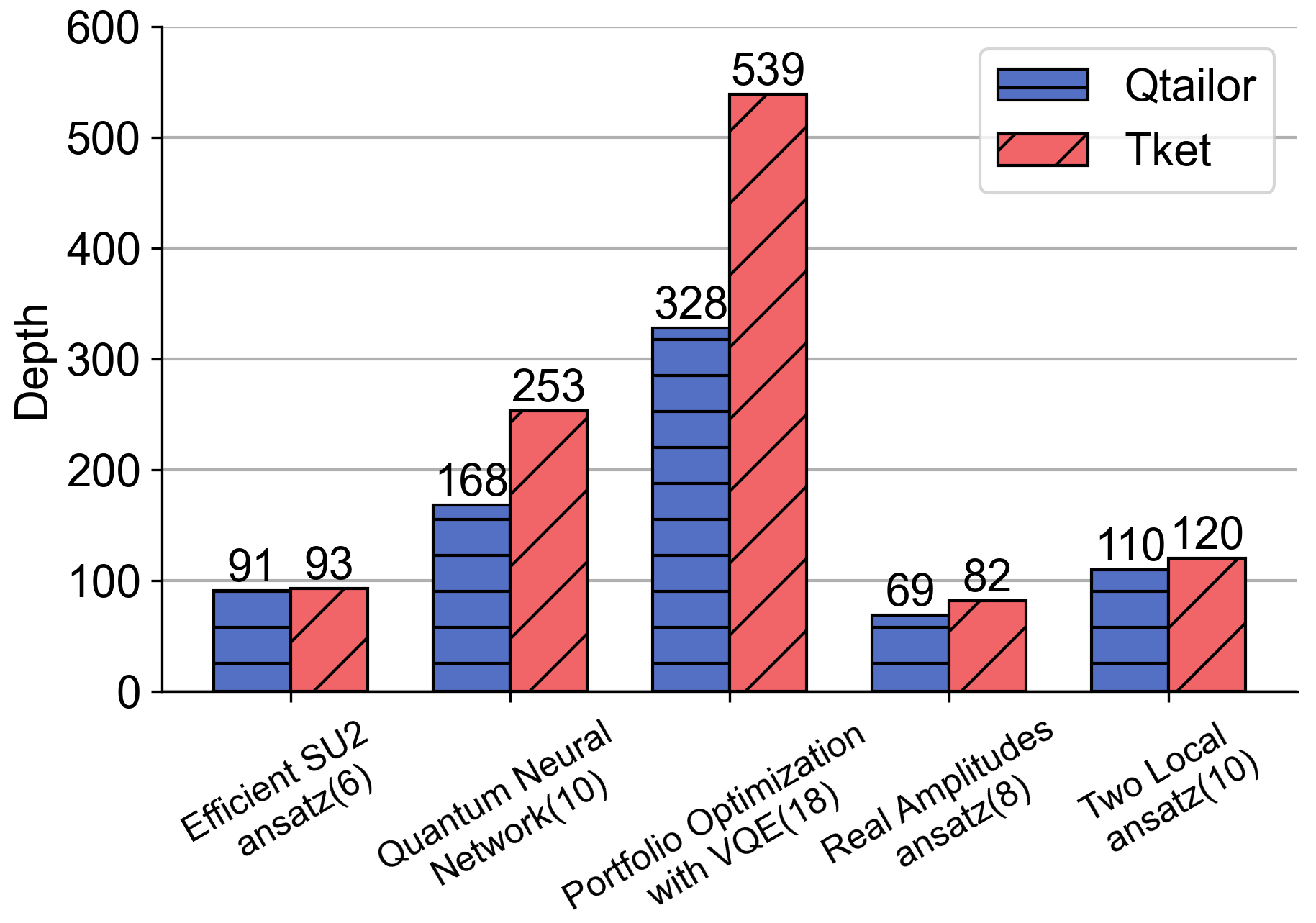}
    \caption{Comparing with the \textit{TKet} compiler, where the horizontal axis represents the types of circuits, with the number of bits indicated in parentheses. } 
    \label{tket_benchmarkBar} 
\end{figure}

\textit{Qtailor} achieves a maximum circuit depth reduction of 39\% compared to Tekt, as illustrated in Figure \ref{tket_benchmarkBar}. The results indicate that employing \textit{Qtailor}'s suggested topology is more effective for reducing circuit depth, and highlights Qtailor's compatibility with various compilers.

\subsection{ Graph Isomorphic: an Example}
\label{app:isomorphic} 
\begin{figure}[tb] \centering
	\includegraphics[width=\columnwidth]{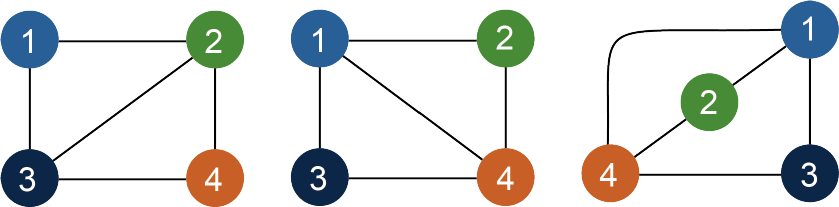}
	\\
	\makebox[0.15\textwidth]{(a)}
	\makebox[0.15\textwidth]{(b)}
	\makebox[0.15\textwidth]{(c)}
	\\
    \caption{A Example of Non-Isomorphism and Isomorphism. Figure (a) and Figure (b) are non-isomorphic because they possess different edges (specifically, edge $ \left \langle 1,3\right \rangle $ and $\left \langle 1,4 \right \rangle$). On the other hand, Figure (b) and Figure (c) are isomorphic. Despite their dissimilar forms, they meet the aforementioned five conditions that have been defined.} 
    \label{isomorphic1}
\end{figure}

Figure \ref{isomorphic1} illustrates the concept of isomorphic and non-isomorphic graphs. Isomorphism refers to the property of a graph where a single graph can exist in multiple forms. In other words, two distinct graphs can possess identical numbers of edges, vertices, and edge connectivity. Such graphs are referred to as isomorphic graphs. For two graphs to be considered isomorphic, they must satisfy the following conditions: (a) Graphs must have the same number of vertices and edges, and their degree sequences must match. (b) If one graph forms a cycle of length $k$ using vertices ${v_1, v_2, v_3, \ldots, v_k}$, the other graph must also form the same cycle of length $k$ using vertices ${v_1, v_2, v_3, \ldots, v_k}$. (c) The adjacent matrices of both the graphs are the same.

\begin{figure}[t] \centering
	\includegraphics[width=\columnwidth]{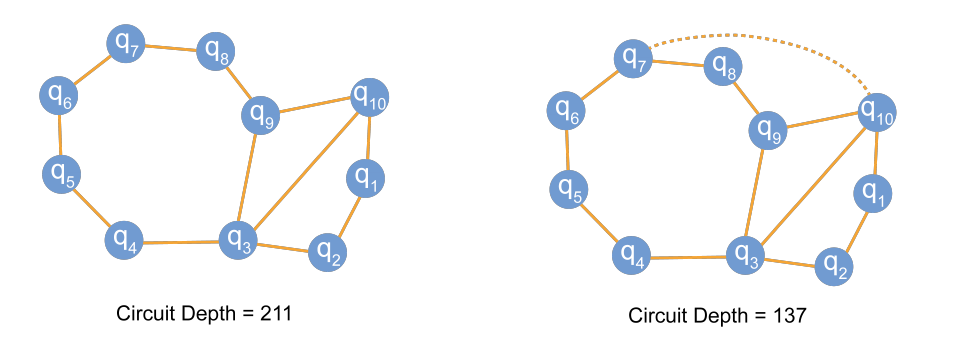}
	\\
	\makebox[0.23\textwidth]{(a)}
	\makebox[0.23\textwidth]{(b)}
	\\
    \caption{Mapped Circuits Depth on Non-Isomorphism Graphs: Figure (b) exhibits an additional edge (represented by a dashed line) in comparison to Figure (a). Upon mapping Quantum Fourier Transformation circuits onto the graph, the depths of the two figures are determined to be 211 and 137, respectively. This signifies a notable reduction in depth of 35.07\% when compared to the former configuration.} 
    \label{isomorphic2}
\end{figure}

In our experiments, we found that the ones that have an effect on the depth of the circuit are non-isomorphic graphs because the non-isomorphic graphs have different adjacency matrices, which means that the connectivity of the nodes in the graphs is different, and Figure \ref{isomorphic2} shows the effect of two non-isomorphic graphs with very small differences on the depth of the circuit.

\subsection{Implementation Details}
\label{app:ImplementationDetails}
\begin{table}[t]\centering
    \label{app:hyperpara}
	\begin{tabular}{llll}
        \toprule
        Hyper-parameters & \textbf{Values}\\
        \midrule
        Critic network &  256, 256 \\
        Policy network & 256, 256 \\
        Discount factor & 0.99 \\
        Batch size & 4000 \\ 
        SGD Minibatch size & 128\\
        Learning rate & 0.00005 \\ 
        Optimizer    & SGD \\
        Use KL loss & True \\
        \bottomrule
    \end{tabular}
    \caption{ Hyper-parameters for RR-PPO }
\end{table}

The benchmark circuits utilized in this study are sourced from the Munich Quantum Toolkit Benchmark Library (MQT Bench) \cite{mqtbench}. These circuits are stored in a .txt file in Qasm \cite{crossOpenQASM2022} format. A brief description of the circuits can be found on the MQT Bench website \cite{circuits_intro}. All experiments were conducted on the Ubuntu 23.04 operating system using an Intel Xeon CPU (3.10GHz). The reinforcement learning model was implemented in Python 3.10 with the RLlib framework\cite{rllib}, which is a scalable and flexible reinforcement learning library built on the Ray distributed execution framework Detailed model hyper-parameters are listed in Table 2. Since Most existing QPU exhibit maximum degree ranging from 2 to 4, represented by structures like linear grids and hexagons, we set the  maximum degree as 4. 

\begin{algorithm}
	\caption{Reward-Replay Proximal Policy Optimization (PPO) Algorithm}
	\begin{algorithmic}[1]
		\REQUIRE Initial policy parameters $\theta_0$, initial value function parameters $\phi_0$
		\FOR  {each $iteration$}
		\STATE  \textcolor{blue}{\textbf{Collect Trajectories(Algorithm 2)}} $\mathcal{T}_{\theta_{old}}$ using policy $\pi_{\theta_{old}}$
		\STATE Compute rewards-to-go $\hat{R}_t$ and advantage estimates
		\FOR {each $epoch$ }
		\FOR{minibatch of size $M$ sampled from $\mathcal{T}_{\theta_{old}}$}
		\STATE Compute ratio $r_t(\theta) = \frac{\pi_\theta(a_t|s_t)}{\pi_{\theta_{old}}(a_t|s_t)}$
		\STATE Compute clipped surrogate objective using ratio $r_t(\theta)$:
		\STATE Update policy parameters $\theta$ by maximizing $L(\theta)$ via stochastic gradient ascent
		\STATE Update value function parameters $\phi$ by minimizing the value function loss:
		\ENDFOR
		\ENDFOR
		\STATE $\theta_{old} \leftarrow \theta$
		\ENDFOR
	\end{algorithmic}
\end{algorithm}

\begin{algorithm}
	\caption{Collect Trajectories with Reward-Repaly}
	\begin{algorithmic}[1]
		\FOR{each episode until batch is full}
		\STATE Reset environment and get initial state $s_0$
		\STATE Initialize trajectory $\tau = \emptyset$
		\WHILE{episode not done}
		\STATE Select action $a_t$ based on  policy $\pi_{\theta_{old}}(a_t|s_t)$
            \IF { The reward $r_{a_t}$ corresponding to the $a_t$ is available in the memory }
            \STATE set  reward $r_t$  = $r_{a_t}$
                \IF { $threshold_{a_t} <=0$ }
                    \STATE clear the pair $\left \langle r_{a_t},a_t> \right \rangle $ in memory
                \ENDIF
            \ELSE
            \STATE Execute the action $a_t$ and run a evaluation on circuit
            \STATE Use the reward function to calculate the reward value and assign it to $r_t$
            \STATE Append $(a_t,r_t)$ to memory
            \STATE Set $threshold_{a_t} $ to  a positive integer
            \ENDIF
        \STATE Update state $s_t \gets s_{t+1}$
		\STATE Append $(s_t, a_t, r_t, s_{t+1})$ to $\tau$
		\ENDWHILE
		\STATE Add trajectory $\tau_{\theta_{old}}$ to the batch of trajectories
		\ENDFOR
	\end{algorithmic}
\end{algorithm}

\subsection{More Details of Force-Directed Grid Layout}\label{appendixForce}
In Section  \ref{sec:layout}, it was previously mentioned that the vertex moves in the direction of the resultant force, which represents the combined effect of all the component forces acting on an object. Component forces refer to the individual forces that contribute to the resultant force.  It is important to note that the vertex moves incrementally, changing its forces in accordance with its position. The new forces are then recalculated, taking into account the small change in position. Subsequently, the vertex moves again based on the updated combined forces. This iterative process is repeated several hundred times until the forces acting on the entire system reach a state of equilibrium.
\begin{figure}[tb] \centering
	\includegraphics[width=\columnwidth]{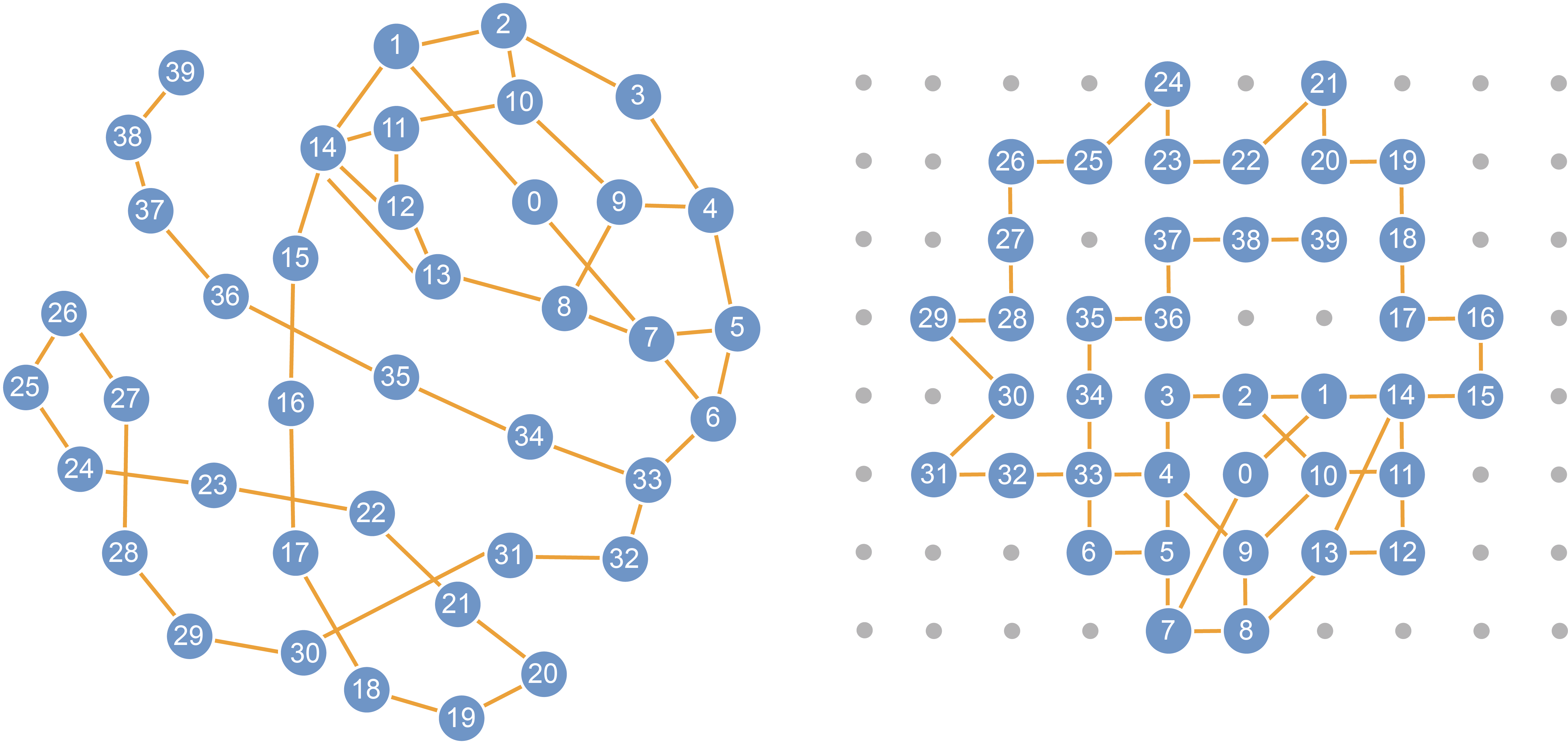}
    \makebox[0.45\columnwidth]{(a)}%
    \hfill
    \makebox[0.55\columnwidth]{(b)}
    \caption{Topology recommend for 40-bit Amplitude\_Estimation circuits by RR-PPO model and  a  force-gird layout for this topology.} 
    \label{ae40topo}
\end{figure}
\begin{figure}[tb] \centering
	\includegraphics[width=0.7\columnwidth]{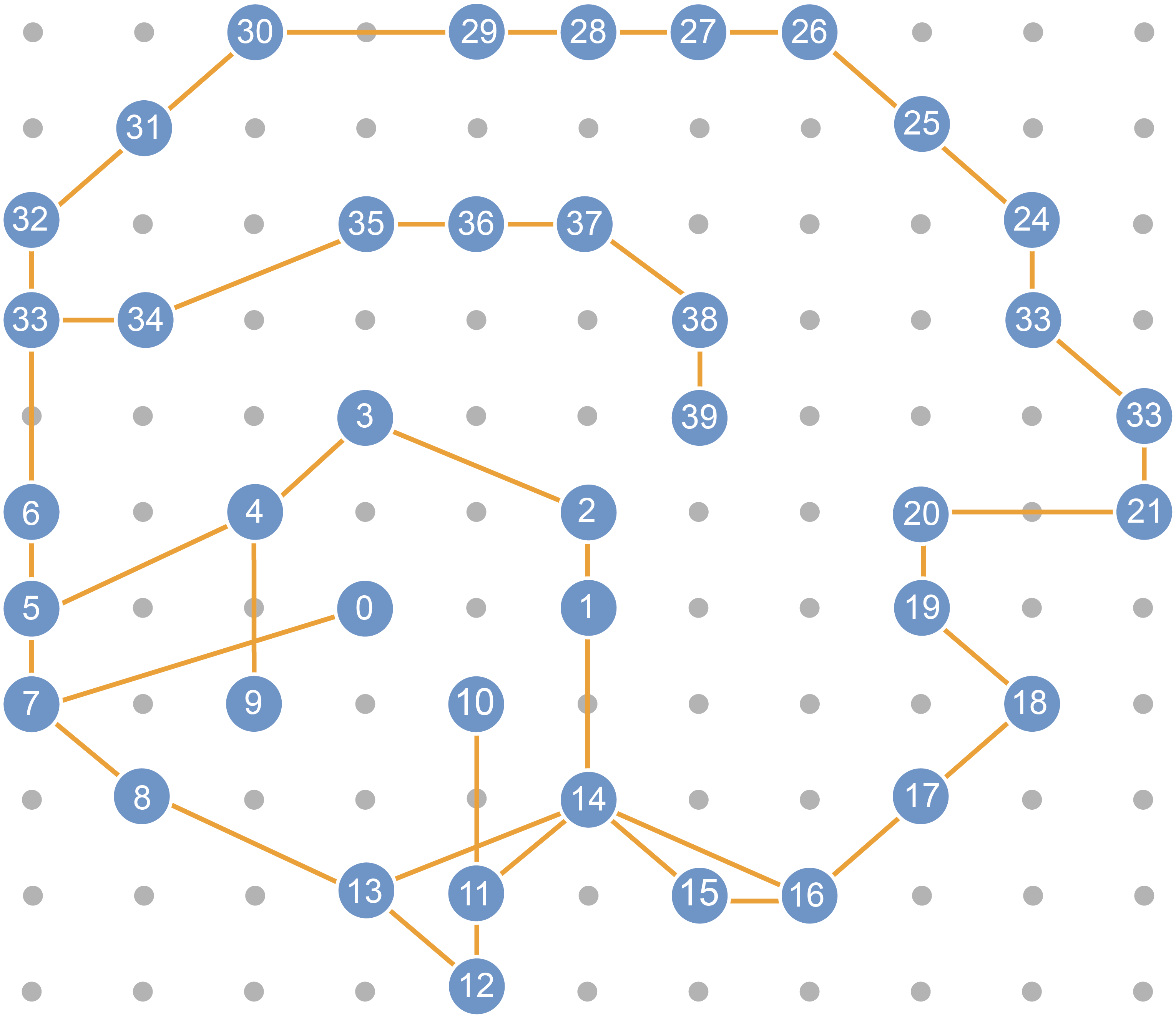}
    \caption{A sparse force-direct gird layout for 40-bit Amplitude\_Estimation circuits} 
    \label{ae40-3} 
\end{figure}
   
\subsection{Recommended Topologies for Circuits}
\label{QtailorRecommended}
In this section, we present the recommended  topology for the Amplitude Estimation circuit with 40 qubits using \textit{Qtailor}. We assign serial numbers to the qubits in the circuit, where the first qubit is denoted as 1. Initially, we illustrate the topology suggested by RR-PPO (refer to Figure \ref{ae40topo}(a)). Subsequently, the qubits are precisely arranged on the grid using the force-grid layout algorithm (refer to Figure \ref{ae40topo}(b)). The force-directed grid layout allows for spacing the qubits more sparsely, reducing potential interference, as shown in Figure \ref{ae40-3}. This can be easily achieved by modifying the coefficient of the repulsion force.
\setcounter{secnumdepth}{0}
\section{Acknowledgments}
H.-L.H. acknowledges support from the National Natural Science Foundation of China (Grant No. 12274464), and Natural Science Foundation of Henan (Grant No. 242300421049).
We gratefully acknowledge Hefei Advanced Computing Center for hardware support with the numerical experiments.
\bigskip
\nobibliography*
\bibliography{aaai25}
\end{document}